\documentclass[sigconf,authorversion]{acmart}

\AtBeginDocument{%
  \providecommand\BibTeX{{%
    \normalfont B\kern-0.5em{\scshape i\kern-0.25em b}\kern-0.8em\TeX}}}

\usepackage{balance}
\usepackage{graphicx}
\usepackage{enumitem}
\usepackage{amsmath,amsfonts}
\usepackage{xcolor}
\usepackage[page]{appendix}

\newenvironment{denseitemize}{
\begin{itemize}[topsep=2pt, partopsep=0pt, leftmargin=1.5em]
  \setlength{\itemsep}{4pt} 
  \setlength{\parskip}{0pt}
  \setlength{\parsep}{0pt}
}{\end{itemize}}

\renewcommand{\paragraph}[1]{\noindent{\textbf{#1:}}}

\newcommand{\sattestor}[1]{\ensuremath{sattestor#1}}

\newenvironment{tightlist}{\begin{list}{$\bullet$}{
  \setlength{\itemsep}{.5mm}
    \setlength{\parsep}{0mm}
}}{\end{list}}

\author{Paul Syverson}
\affiliation{%
  \institution{U.S. Naval Research Laboratory}
  \country{}
}
  \email{paul.syverson@nrl.navy.mil}

\author{Matthew Finkel}
\affiliation{%
  \institution{The Tor Project}
  \country{}
}
  \email{sysrqb@torproject.org}

\author{Saba Eskandarian}
\affiliation{%
  \institution{UNC Chapel Hill}
  \country{}
}
  \email{saba@cs.unc.edu}

\author{Dan Boneh}
\affiliation{%
  \institution{Stanford University}
  \country{}
}
  \email{dabo@cs.stanford.edu}

\title{Attacks on Onion Discovery and Remedies via
Self-Authenticating Traditional Addresses}

\copyrightyear{02021} 
\acmYear{02021} 
\setcopyright{acmlicensed}\acmConference[WPES '21]{Proceedings of the 20th Workshop on Privacy in the Electronic Society}{Nov.\ 15, 02021}{Virtual Event, Republic of Korea}
\acmBooktitle{Proceedings of the 20th Workshop on Privacy in the Electronic Society (WPES '21), November 15, 02021, Virtual Event, Republic of Korea}
\acmPrice{15.00}
\acmDOI{10.1145/3463676.3485610}
\acmISBN{978-1-4503-8527-5/21/11}

\begin{document}



\begin{abstract}
  Onion addresses encode their own public key. They are thus
  self-authenticating, one of the security and privacy advantages of
  onion services, which are typically accessed via Tor Browser.
  Because of the mostly random-looking appearance of onion addresses,
  a number of onion discovery mechanisms have been created to permit
  routing to an onion address associated with a more meaningful URL,
  such as a registered domain name.

  We describe novel vulnerabilities engendered by onion discovery
  mechanisms recently introduced by Tor Browser that facilitate hijack
  and tracking of user connections. We also recall previously known
  hijack and tracking vulnerabilities engendered by use of alternative
  services that are facilitated and rendered harder to detect if the
  alternative service is at an onion address.

  Self-authenticating
  traditional addresses (SATAs) are valid DNS addresses or URLs that
  also contain a commitment to an onion public key. We describe how
  the use of SATAs in onion discovery counters these vulnerabilities.
  SATAs also expand the value of onion discovery by facilitating
  self-authenticated access from browsers that do not connect to
  services via the Tor network.

\end{abstract}



\maketitle

\section{Intro to Onion Addresses and SATAs}
\label{sec:intro}

Our presentation of the attacks and countermeasures that are the
subject of this paper assumes a familiarity with SATAs. The
extremely brief description of these in this introduction is supported
by a more detailed description in Appendix~\ref{sec:appendix}.
But first we give a similarly brief description of onion addresses.

Tor's onion addresses are self-authenticating: the address encodes a
public key used to authenticate the address. Onion addresses
are an IETF standard~\cite{ietf-onion-tld-rfc} used by many Fortune
500 companies, government agencies, and major media and news
organizations. They are generally only reachable via the Tor network,
typically via Tor Browser. The Tor Project's overview of onion
services~\cite{onion-service-overview} provides a basic description, a
list of some notable onion sites, and links to further documentation.
Though self-authenticating, onion addresses are comprised only of an
encoding of an ed25519 public key (plus a checksum and version
number).  They thus generally appear to be meaningless, random-looking
56-character strings, though some large sites commit significant
computational resources to make a portion of the address meaningful,
e.g., \texttt{facebookwkhpiln\\
  emxj7asaniu7vnjjbiltxjqhye3mhbshg7kx5tfyd.onion}.

Self-Authenticating Traditional Addresses (SATAs), embed security into
the nodes of the web in the same way that onion addresses do.  Unlike
onion addresses, they are also traditional addresses understood by
DNS, by humans (to the extent registered domain names are), and by
popular browsers that know only the protocols and standards of the
(less-secure) web. A working example described
in Appendix~\ref{sec:appendix} is \vspace*{-.4em}
\begin{quote}
\texttt{https://selfauth.site/?onion=ixxuq4b4bsr3agg
  bokovydiiys7rolq4ewqjva67qfpmp3y55jsxi5yd}.
\end{quote}
\vspace*{-.4em}

Connecting to a SATA is authenticated by the traditional mechanisms of
TLS and TLS certificates, but to support autonomy of site owners, it is
also authenticated by a credential that attests to the binding of the
domain name, the onion address, and optionally contextual labels about
the type of site, e.g., that it is a news media site, or is a domain
owned by Microsoft. We call such an attestation a
\emph{sattestation}. Such sattestations are made by a SATA, either
a third-party SATA, or by a SATA about itself, in which case it also
includes a fingerprint of the TLS certificate. 

Currently a WebExtension we have implemented for Tor Browser and
Firefox can recognize SATAs and checks sattestations. Ultimately we
expect this functionality to be in the browser itself, but a WebExt
facilitates initial deployment, experimentation, and development.  A
SATA website will create a self-sattestation, which is a signature
using the onion private key over the following: 
the domain name and onion address, a fingerprint of the TLS cert, an
issue date and most recent recheck date, and
(optionally) a list of contextual labels. After completing the TLS
handshake, a SATA-aware browser checks the self-sattestation and
either succeeds and continues with the connection or returns an error
and warning.

In Section~\ref{sec:airdrop} we describe various
mechanisms for onion discovery that support routing to an onion
address associated with a more meaningful URL\@. We also set out
vulnerabilities that these mechanisms create or facilitate. Finally,
in Section~\ref{sec:remediation} we describe how SATAs and
sattestation remediate these vulnerabilities.

\section{Attacking onion service discovery}
\label{sec:airdrop}

Tor recently added mechanisms to make it easy for a 
Tor Browser client that connects to a registered domain 
to be easily (and mostly automatically) routed
and connected to an associated onion site instead, when one is available.
This follows a well-established tradition of automatically choosing a more
secure option for the user, when one is available.
HTTP Strict Transport Security (HSTS) is a classic example of 
such an automated choice.
These automated Tor mechanisms are deployed and integrated into existing stable
releases of Tor Browser. In this section we will describe novel
attacks resulting from these Tor security deployments. 

\medskip
\paragraph{1. Onion Alternative Services} 
Alternative Services (RFC~7838) present via an HTTP
header an alternative (alt) URL to which a browser should connect
instead of the one selected and displayed in the URL bar. This is
entirely invisible to the user unless she examines the source code for
this header. Cloudflare began offering onion alt services for
its supported domains in April 02018\footnote{This paper uses 5-digit
  year dates in support of long-term thinking and awareness. Cf.\
  https://longnow.org/.}; Tor users could request
connection to an HTTPS URL but be routed and connected to the domain
through its onion address~\cite{cloudflare-alt-onion-blog}.  The URL
address bar shows the requested HTTPS URL, even though the connection
is through the onion address.  No changes were needed to Tor Browser
for these to function as intended.

An attacker who temporarily obtains a valid certificate for a victim
domain can redirect browsers visiting (a spoof of) the victim domain
to an alt service at the attacker's onion address.
Because browsers cache the alternate service response header for a period
of time, all subsequent requests to the victim domain will go directly
to the attacker's onion site, without ever connecting to the victim website.
Neither the user nor the victim website will detect this.

This vulnerability was previously discussed in~\cite{secdev19}.
As we will see in the next section, 
it can be elegantly mitigated using SAT addresses. 
As an aside, we point out that the caching
behavior of alt services enables difficult to detect user tracking.

\medskip
\paragraph{2. Onion location}
Tor Browser 9.5 (released June 02020) introduced
support for \emph{onion location}.
Owners of websites at traditional domains 
can provide a header over HTTPS that will cause a ``.onion
available'' prompt to display in the URL bar. Clicking on this will
allow redirection to an onion address,
amounting to a kind of attestation by the site owner to its onion address.
However, in case an attacker has a fraudulent or hijacked TLS certificate
for some domain, they can set up a hijack site with onion location
pointing to an onion site run by the attacker.

Tor Browser also allows users to click a radio button in their
settings to always ``Prioritize .onion sites when known''. (The
default is ``Ask every time''.) A client receiving any
onion-location header (or onion-location meta tag) thus
automatically redirects to the indicated onion address with no user
input. This makes typo-squatting, domain impersonation, and similar
attacks more likely to escape detection: when clicking on a link, the
initial URL is visible for only a second before redirection to a
random-looking onion address.   

For example, suppose an adversary sets up a link:\\
\texttt{<a href=''https://fuii.com''>full.com</a>} where the
adversary's domain \texttt{https://fuii.com} sends an onion location
redirect to an onion address controlled by the attacker.  Then a user
would need to (a)~be looking at the URL bar during the second that
\texttt{fuii.com} rather than \texttt{full.com} was displayed and
(b)~notice and respond to the difference, or (c)~recognize
arrival at the wrong onion address. (We have deliberately
not chosen examples using even more visually similar URLs, e.g.,
using unicode, so the reader could see the point being made.)
Further, domain impersonations will leave no relevant trace in CT
logs~\cite{domain-impersonation-acmccs2019}.  Onion location also
dovetails nicely with targeted attacks, e.g., with links on a
particular site that is primarily visited by a unique class of users
or on a page offered to specific logged-in or otherwise identified
users.

It might seem that an adversary who controls a server enough to set an
onion location redirect could simply set an ordinary HTTP location
redirect to the onion service, so there is no real increase in
vulnerability from making onion location available. But just as any
access by non-Tor-aware browsers would get a confusing failure from a
nonmalicious HTTP redirect to an onion address, this would also make
such a redirect in an attack highly visible.  So onion location is 
increasing
vulnerability for attackers not wishing to loudly announce the hijack
they are conducting. The adversary could add a redirect back to
\texttt{full.com} for any connections reaching \texttt{fuii.com} from
a non-Tor-exit IP to reduce that, as well as limiting posting of the
pharming link to pages on onion services. Even so, Tor Browser users
may be surprised and suspicious from the automatic redirect to an
onion address. All of which means that onion location does
significantly increase the attack possibilities. Fortunately,
the remedies in Section~\ref{sec:remediation} address the potential
attacks whichever type of header causes the redirect.

\medskip
\paragraph{3. HTTPS Everywhere for onion addresses}
HTTPS Everywhere (H-E) is a browser extension originally intended to
replace attempts to connect to a website via unencrypted HTTP with a
TLS-encrypted connection to the same
website~\cite{https-everywhere}. Since some sites used a different URL
for the secure version of the same webpage, or put different content
at the same URL when an HTTPS connection was used, it was not enough to
simply rewrite a request for\\ \texttt{http://example.com} to
\texttt{https://example.com}. 
To address this, HTTPS Everywhere created a ruleset that the
extension used to make the appropriate rewrite of URLs in the ruleset.

In 02016, Syverson and Boyce~\cite{ieeesp-pgp-onion}
suggested using an HTTPS Everywhere
ruleset as a way for users to type in a familiar, understandable domain
name in an address bar that would automatically be converted 
to a connection to an onion address. 
Subsequent work noted that SATAs
could address the problem of a user experiencing a
surprising URL change to a completely different looking URL from the one they
entered~\cite{once-and-future} (similar to onion location above),
and implemented integration of
H-E's address rewrites into a WebExtension for authenticating and
validating SATAs~\cite{secdev19}.

Originally, there was a single H-E ruleset, updated whenever periodic
software updates were made for the H-E WebExtension. In
02018, H-E introduced ``update
channels''~\cite{HE-continual-ruleset}. This permitted ``continual
ruleset updates'' without a need to update the browser extension
itself. This also permitted others besides the H-E maintainers to
create and maintain authenticated ruleset update channels.

\medskip
\paragraph{H-E Rules for SecureDrop}
In cooperation with Freedom of the Press Foundation (FPF), Tor has recently
begun providing a means for users to type, e.g.,
\texttt{www.cbc.ca.securedrop.tor.onion} in the URL bar of Tor Browser
and be connected to
\texttt{gppg43zz5d2yfuom3y\\ fmxnnokn3zj4mekt55onlng3zs653aty4fio6
qd.onion},
the onion address of the SecureDrop instance run by the Canadian
Broadcasting Company\@. A primary component enabling this is an H-E ruleset
channel maintained and signed by FPF~\cite{securedrop-ruleset}.
Signed update channels are similar to sattestation:
someone the client trusts signs a set
of understandable names based on registered domains and, for
each of those, an associated onion address. 

Leveraging this update channel, Tor Browser 
recognizes addresses of this form, and continues to
display the familiar understandable URL though the connection is made
to the associated onion address using the onion service lookup and
routing protocols. This addresses the ``surprising URL change'' issue
noted above. 

\medskip
\paragraph{The problem}
Unfortunately the FPF H-E channel and described Tor Browser
functionality are quite vulnerable to a compromised or malicious 
creator of a trusted channel.
Such a creator can assign any onion address to
\texttt{example.com.securedrop.tor.onion}. 
For H-E rulesets there is no specific infrastructure, like certificate
transparency~\cite{certificate-transparency-site}, to detect malicious
or conflicting assignments of onion addresses to a domain name. Nor is
there any current design or implementation that uses existing
infrastructure for such detection.

To make matters worse, 
a rewrite from a ruleset takes place entirely in the browser
before any external communication.
Such an attack can thus succeed without any need to
also succeed at a DNS hijack or to obtain a rogue TLS certificate.  
Further, a hijacked rewrite guarantees a successful attack:
connection attempts \emph{never} reach the correct destination during
an H-E ruleset based attack.

Moreover, if Alice receives from someone she trusts at the CBC
the address \texttt{www.cbc.ca.securedrop.tor.onion}, she must
separately trust that the H-E channel maintained by FPF is mapping
this address to the proper onion address. Further, the default URL displayed
in Tor Browser is \texttt{www.cbc.ca.securedrop.tor.onion}. So, even
if she knows the correct associated onion address, unless she does
some separate additional checking, she will not have any indicators of
(in)consistency with the address displayed to her. Perhaps more
importantly, neither will her software. 
Hiding from the browser the mapping of
recognizable names to onion addresses makes it harder to counter
any unicode, doppelganger, or typo-squatting attacks on the ruleset.

We now turn to the use of SATAs and sattestation to counter 
the attacks described in this section.

\section{Attack remediation}
\label{sec:remediation}

Onion location, discussed in the previous section,
 was introduced to provide a simple way for Tor
Browser users to connect to a more secure onion site for the current
webpage when one is available. 
Note that if onion location redirects to a SATA rather than to a
simple onion address, it would make the self-authentication
protections of onion addresses available to users of every
browser that has our extension installed, not just Tor Browser.
Before describing how SATAs better secure onion location, however, we
recall a previously published mitigation that SATAs provide to
vulnerabilities facilitated by onion alt services.

\subsection{Remedies via SATA}

\paragraph{Trusted alt services}
Onion alt services have been discussed in the context of SATAs
in previously published work~\cite{secdev19}. We do not revisit the
full discussion here and simply recall that (1) neither the user nor
the browser itself can readily tell whether the connection is actually
to an onion address or not (so not actually self-authenticating in a
practical sense), and (2) onion alt services provide easy to implement
and hard to notice tracking of users. As described in~\cite{secdev19},
however, the WebExtension, keeps track of alt-svc headers it sees, indicates
if there is a self-sattestation provided for an onion alt service, and also 
permits blocking of alt services unless they are trusted.
This provides a basic interface for user visibility and control over
the alternative services accepted.

\medskip
\paragraph{Self-authenticating onion location} 
if onion location always redirects to a SAT address,
then the redirected connection can still be an HTTPS connection
using the same TLS certificate as the original connection
(recall that the SATA is included in the TLS certificate for the domain).
This has two advantages. 

First, there is no switch to HTTP for the onion connection. 
This is primarily a UI rather than a security benefit 
(onion connections are always encrypted).
It ensures that the browser does not show any HTTP warnings.

Second, since the onion address is incorporated in the certificate,
a CT log will have promised to include it using
a signed certificate timestamp (SCT). Neither Firefox
nor Tor Browser natively supports checking SCTs at time of writing.
But even before that changes, e.g., as described in~\cite{ctor-popets},
SATAs are accessible to other popular browsers that \emph{do}
check SCTs. Using them thus raises the bar for attacks
intended to avoid such checks.

With SATAs, onion location as a separate feature
can be abandoned altogether 
and we can still have both of these advantages. 
If a site is provided as a SATA, a ``.onion
available'' prompt is not needed. 
If a requested URL is available as a SATA, 
then the browser should be redirected to the SATA automatically. 
Since both the registered domain
and the onion address of the SATA are represented in the TLS
certificate, the browser can safely reach the requested
site over HTTPS using the SATA\@. 
And, as we have newly implemented,
if using Tor Browser the WebExtension will now automatically reroute
to the onion site without touching DNS or otherwise exiting the Tor network.
The URL bar will, nonetheless, continue to display the SATA rather than
switching the displayed URL to the onion address.
Since SATA use for improved authentication does not automatically
imply a site is generally accessible via onion service protocols, it
would make sense to add a field or bit to the SATA header indicating
whether or not it is onion-service accessible.

Another advantage of SATAs over onion location
is that, since a SATA is a fully qualified domain name
(FQDN), the destination can have a free domain-validated (DV) certificate
right now, rather than depending on Let's Encrypt to implement cert issuance
of DV certificates for onion addresses. HARICA does issue DV certificates
for onion addresses. Since issuance involves both the usual checks for
control of the registered domain and possession of the private onion key,
this amounts to an implicit structural sattestation from HARICA\@.
In addition, using SATAs, site owners do not
need to set up and maintain two versions of the site, one with
addresses based on the registered domain name and one with onion
addresses.

\paragraph{Reflecting domain authentication decisions in CT logs}
Onion location does not incorporate into CT logs
any association between the traditional domain and the onion
address. CT logs are primarily meant to provide a record of
certificate authority (mis)behavior, but more generally they provide a
public record of the certificates that underlie authentication on the
web. Onion location significantly affects that
authentication but does not have any associated components that change
this public record in any way. Thus, it not only engenders the
attacks we describe but also this secondary vulnerability. But, in
addition to countering the attacks those mechanisms engender, SATAs
are reflected in certificates and CT logs in ways that provide
relevant evidence if attacks do occur.
Third-party CT log inspectors exist that site owners can
use to check for the existence of certificates they did not authorize.
Adding a check for any claimed association of their onion addresses
with other registered domain names should be straightforward. These
would not fool the WebExt or Tor Browser since the attacker would not have
the onion private key, but it would reveal attempts to use their
SATAs and onion addresses to attack less-secure browers in some way.

\paragraph{Self-authenticating SecureDrop}
Like onion location, the FPF SecureDrop H-E ruleset is about discovery
rather than authentication, and it enables attacks
on authentication if the ruleset is controlled by the adversary.
For example, suppose a user connects to
\texttt{https://www.cbc.ca/} and finds on that page a link to\\
\texttt{http://www.cbc.ca.securedrop.tor.onion}.
Tor Browser converts this to an onion address using the H-E ruleset channel 
maintained by FPF, and then connects to that onion address.
Thus, if the FPF ruleset is compromised, the user will end up at the wrong 
SecureDrop location. 

Addresses like \texttt{www.cbc.ca.securedrop.tor.onion} are
human-meaningful extensions of traditional, registered domains.  They
are easy enough even to type by hand, given a known domain name.
Unfortunately, they are not proper domain names themselves, so cannot
have TLS certs. Fortunately, due to implementation decisions, when
routing to an onion address based on a SecureDrop H-E rule, Tor Browser
performs TLS certificate validation based on the underlying onion
address, instead of the human-meaningful name. And, the
onion address for the CBC SecureDrop site,\\ \texttt{[onion].onion},
can be included as an
alt name in the TLS cert for \texttt{www.cbc.ca}, as some journalism
sites have done. The cert could also contain as alt name the SATA
\texttt{[onion].www.cbc.ca}. Thus, if the
URL bar contains\\
\texttt{www.cbc.ca.securedrop.tor.onion/?onion=[onion]},
e.g., as\\ linked from \texttt{www.cbc.ca}, then the SecureDrop 
rewrite can induce a connection to [onion].onion, and the 
WebExtension can then check the returned TLS cert and SATA header
against the result of removing \texttt{.securedrop.tor.onion}
from\\ \texttt{www.cbc.ca.securedrop.tor.onion/?onion=[onion]}.

\subsection{Remedies via trusted sattestors}

Given the ability to hijack a certificate, an adversary can obtain a
TLS cert for some SATA for the registered domain, one that she
can hijack \emph{and} authenticate to a SATA WebExtension---because
she has herself created the onion address component for the
certificate. This is what sattestation counters. (If the
certificate contains the onion address as a SAN and the CA checks
possession of the private onion key, as HARICA currently does, then an
external attacker cannot hijack a SATA's certificate issuance. A
malicious CA still could.) If an extension is configured to require
sattestation from trusted sattestors from which the adversary is not
able to obtain a sattestation, then such a hijack against onion
location (or against onion alt services or the SecureDrop H-E ruleset)
will not succeed.

Browsers can be configured to expect third-party sattestations with
corresponding labels for specific domains.  For example, large
enterprises can cooperate with browser vendors so that labels for
their domains with SATAs are correct. To limit the need for specific
enterprise/browser interactions, industry consortia (e.g., software
alliance BSA) can further provide sattestations that indicate
sattestors and labels for member entities. If a sattestor SATA and
label for a domain is on this list, the specific SATA for the domain
need not be on it: the browser will expect, e.g., some SATA for
\texttt{live.com} with label \emph{microsoft} and a sattestation from
the BSA that labels a Microsoft SATA with
\mbox{\emph{sattestor(microsoft)}}.  In this way the only SATA that
must be trusted in advance for any consortium member is that of the
consortium.

The existing WebExt does not support automated trust decisions
for onion alt services. Contextual sattestations could underly
such automation, however, which would counter third-party hijack via
alt services but not first-party tracking (without
additional checks).

Sattestation uses the same
mechanism for trust at the scale of personal associations as for trust
at the scale of a large corporation or a certificate authority.
Clients are assumed to reflect contextual trust that users attach to
specific domain names or to sattestation labels. For sites that have
no such context, sattestation provides no more trust than a TLS
certificate issued by a generic CA\@.  But, clients can be
configured with some default contextual labels and trusted sattestors
for those labels. They can then trust the binding of a SATA with a
particular label for a domain, e.g., journalism, even if they are
completely unfamiliar with the domain.

\section{Conclusion}
\label{sec:conclusion}

We have described novel vulnerabilities engendered by onion discovery
mechanisms recently introduced by Tor Browser that facilitate hijack
and tracking of user connections as well as previously known hijack
and tracking vulnerabilities. We described how the use of SATAs and
sattestation in
onion discovery counters these vulnerabilities.  SATAs also expand the
value of onion discovery by facilitating self-authenticated access
from browsers that do not connect to services via the Tor network and
by making onion addresses visible in CT logs and making SCT checks of
associated certificates visible in other browsers.

\section*{Acknowledgments}
The authors would like to thank Rasmus Dahlberg for helpful
discussions and comments.
This work was funded in part by ONR.

\appendix
\section{Appendix: SATAs and Sattestation}
\label{sec:appendix}

This appendix provides an overview of SATAs and sattestation.  Though
many of the core concepts are revisited from~\cite{secdev19}, we also
describe novel content introduced here. This includes the
query-string format for SATAs, sattestation credentials and
contextual labels, and an improved implementation including both
these and Tor Browser connections to SATAs via onion service protocols.

Self-Authenticating Traditional addresses, introduced
in~\cite{once-and-future,secdev19}, are designed to enhance
traditional certificate based authentication widely used on the web.
In particular, they reduce the reliance on certificate authorities
(CAs) to provide domain validation by properly binding a DNS domain
name to a public key.

A simple example of a SATA is a DNS domain name where
the site's public TLS key is included as a subdomain.
A site \texttt{bank.com} could advertise its domain
name as \texttt{[HashedPubTLSKey].bank.com}, for example.
A user who searches for the bank on a trusted search engine, 
or bookmarks the bank's homepage,
would reach the bank through this domain name.
A SATA-aware browser will verify the bank's certificate, 
as browsers do today,
and will also check that the public key in the bank's
certificate matches the \texttt{[HashedPubTLSKey]} in the subdomain.
We stress that SAT addresses are intended to enhance 
the current PKI trust model, not supplant it.  

This hashed-TLS-certificate-public-key approach 
is unworkable for a site that has multiple certificates
for a single domain, and it can be inconvenient
with regular certificate rotation if keys
are changed upon obtaining a new certificate.
In Section~\ref{sec:format} we will consider
other options for creating SATAs.
Most notably we will use Tor onion addresses~\cite{ng-onion-services-224}
to mitigate the shortcomings of the 
above approach.
An onion address contains the public key of the 
onion service to which it belongs.
Therefore,\\ $\texttt{[onion-address].bank.com}$
is a viable way to form a SAT address
for \texttt{bank.com}, as explained in Section~\ref{sec:format}. 
As we will see, a single SAT address can be used
to authenticate many TLS certificates. 
Note that here we focus on the
authentication that onion addresses provide,
rather than privacy.

This paper introduces a number of new techniques to make SATAs
compatible with the modern web and demonstrates that adoption of SATAs
can solve security issues in deployed systems,
supporting a traditional trusted internet connection
(TIC)~\cite{cisa-tic-30-trad} use case, simplify certificate
revocation, and strengthen the web's PKI, all while maintaining
backwards compatibility and interoperability with existing
infrastructure.  We summarize these contributions in more detail below
after providing necessary background on SATAs.

Useful and easy to deploy SATAs must satisfy three properties:
\begin{denseitemize}
\item \emph{Authority-independence:}
Once a domain owner creates a SATA for its domain,
a misbehaving CA cannot bind a rogue public key 
to that SATA.
The SATA binds the domain to its public key. 

\item \emph{Dirt Simple Trust:}
Learning a SATA
from a trusted party suffices to ground assurance
that a connection is to the intended site. 

\item \emph{Synergistic Backwards Compatibility:} 
SATAs should be fully backwards compatible with 
existing internet and web infrastructure,
without having to duplicate content or links on sites. 
\end{denseitemize}

SATAs are most useful when the user has a way to obtain 
the correct address for the domain that she wants to interact with. 
This could be through a trusted search engine, a friend,
a previously created bookmark, etc. The existing web PKI also functions
this way:
a user needs to know the full name of the domain that she wants to connect to,
otherwise the subject name validation in an X.509 certificate is meaningless. 
To illustrate this point, consider the secure operating system
\emph{Qubes OS} which is served from the domain {\tt qubes-os.org}.
Users looking to download the operating system often find its domain 
via a search engine query.
If instead a user guesses the address and
connects to {\tt qubes.com} or {\tt qubesos.com},
she will end up at a site that has a valid X.509 certificate,
but is unrelated to the qubes-os project.
That site could serve a backdoored version of the OS,
and the user will never know. 
The same applies to banking sites, shopping sites, etc.

Self-authentication does not automatically guarantee
an intended address either. {\tt
  sik5nlgfc5qylnnsr57qrbm64zbdx6t4lreyhpon3y} {\tt chmxmiem7tioad.onion} is
a self-authenticating address that encodes its own public key, 
but unless told by a trusted source that this is the
onion address associated with {\tt qubes-os.org}, which it is,
the user could not tell that this address is correct. 
The trusted source can attest to this association by signing a
statement for the user asserting a binding of these addresses.  We
describe a means to effectively do that, \emph{sattestation}, in
Section~\ref{sec:sattestation}.
We will also see in Section~\ref{sec:format} that the query-string format for
SATAs introduced in this paper allows a response of\\
\texttt{https://bank.com/?onion=[onion-address]}
to a request for\\ \texttt{https://bank.com}.
A client can 
learn about a SATA simply by being redirected to it from its parent
domain, but the query string format allows this to happen without
any change to the domain name in the URL bar.

\medskip
\paragraph{The SATA threat model}
SATAs are designed to defend against an
adversary whose goal is to fool the user into connecting
to a fraudulent copy of a victim site (e.g., a copy of {\sf bank.com}) 
without the user's knowledge.
To do so, the adversary can do the following:
\begin{denseitemize}
\item \emph{Network attacks:}
the adversary can manipulate DNS for a targeted class of users,
and can potentially mount a BGP hijack of traffic to an IP
range covering an intended destination.
In particular, this ability enables the attacker to obtain a certificate
for a victim site from some existing CAs~\cite{birge2018,cert-hijack-hotpets17}.

\item \emph{CA compromise:}
We explicitly give the adversary the ability to obtain a valid TLS certificate
for the victim site from a reputable CA.
However, the adversary cannot compromise the certificate transparency (CT)
infrastructure~\cite{certificate-transparency-site}, 
and in particular, cannot forge a signed certificate timestamp (SCT).
\end{denseitemize}
SATAs are designed to \emph{fail-closed} in the presence of such adversaries.
See Section~\ref{sec:auth} and~\cite{secdev19} for additional
details on this point.

\medskip
\paragraph{Ghost domains}
To demonstrate the potential benefits of SATAs, we begin with an example: ghost domains.
Consider a web page at {\tt a.com} that points to a URL at {\tt b.com}.
If the domain {\tt b.com} is later abandoned by its owner,
and is taken over by someone else, 
then a user visiting {\tt a.com} will be fooled into going to
the wrong page.  
This is a common occurrence on the web. 
For example, in 02015, bottles of Heinz ketchup 
in Germany carried a label that included a QR code pointing to 
a domain for a recently expired promotion, 
\texttt{sagsmitheinz.de}, 
(translation, ``say it with Heinz''). 
While bottles were still in circulation, this domain
registration expired, and a German pornography company registered
the name, surprising some ketchup customers who followed
the address on their bottles~\cite{sagsmitheinz}. 
Similarly, abandoned banking domains 
have been used to target users who attempted
to connect to those banks~\cite{banking-ghosts}.
More recent research has shown how stale NS records
for an active domain can serve as reactivated zombies
in a resolver's cache~\cite{zombie-ccs20}.

SATAs are a solution to ghost domains.
If a different entity takes over an expired domain name,
that entity will not be able to authenticate as the 
associated SATA to any browser that
checks knowledge of the private key associated with the SATA\@. 
In the ketchup example above, users who try to connect to the expired 
SATA domain in the QR code will get an error.
Likewise for a server incorrectly reached through an attack on DNS or BGP. 
Of course, a domain should be allowed to rotate its SAT address.
This can be done by the domain publishing a new SAT address,
and is further simplified by the sattestation process
we describe in Section~\ref{sec:sattestation}.

\subsection{SAT Addresses in Practice}

\subsubsection{Backwards Compatible SAT Addresses}
\label{sec:format}

Previous work on SAT addresses~\cite{once-and-future,secdev19}
proposed the subdomain format discussed in the previous section, but
using an onion address for authentication.
The SAT address for \texttt{bank.com} 
is then \texttt{[onion-address].bank.com}, 
where ``[onion-address]'' represents an onion address in the format
used by Tor onion services~\cite{ng-onion-services-224}. 
Recall that this is 56 characters long and comprised of a base-32
encoding of the tuple 
(ed25519 public key, checksum, version number, ``.onion'').   
A browser that understands SAT addresses,
such as one using our WebExtension,
will authenticate the TLS certificate for \texttt{bank.com} 
using the onion address, as explained in Section~\ref{sec:auth} below.
(Implementation in an extension allows for both easy experimentation
and configuration on initial rollout, and it facilitates quicker
initial adoption. Once the value add is well-established, the
functionality provided by the extension can be moved into the browser
itself.) If the browser was given the correct SAT address, 
then the user is assured that the connection 
is to the correct \texttt{bank.com},
even in the presence of malicious CAs or ghost domains.
As an example, we set up the following SATA
for the domain \texttt{selfauth.site}:
\vspace*{-.4em}
\begin{quote}
\texttt{https://ixxuq4b4bsr3aggbokovydiiys7rolq4ew}\\
  \texttt{qjva67qfpmp3y55jsxi5ydonion.selfauth.site}.
\end{quote}
\vspace*{-.4em}
This entire string can be used as a
Subject Alternative Name (SAN) 
in a domain validation (DV) certificate issued for 
\texttt{selfauth.site}. This means that a no-cost certificate
for this SATA can be obtained from Let's Encrypt through 
a relatively simple automated process.

These self-authenticating traditional domains 
are the only form of SATA set out in prior work.
This format is fully backwards compatible: 
browsers that know nothing about SAT addresses 
will handle TLS authentication as they do for any other site. 
In this paper we introduce another SATA format with additional
advantages.



\medskip
\paragraph{SAT addresses using query strings}
To be a SATA, the onion addresses need not appear in the domain itself.
We can put the onion address elsewhere too, e.g., in a query
string.  Using a query-string-formatted SAT address, a link to the
above address becomes \vspace*{-.4em}
\begin{quote}
\texttt{https://selfauth.site/?onion=ixxuq4b4bsr3agg
  bokovydiiys7rolq4ewqjva67qfpmp3y55jsxi5yd}.
\end{quote}
\vspace*{-.4em}
Our WebExtension recognizes this SATA format
and uses the SATA to authenticate the 
TLS certificate, as in Section~\ref{sec:auth}.
A SATA-unaware browser will connect to \texttt{selfauth.site}
and authenticate it using only its TLS certificate.

One benefit of this SATA format is that it is not necessary to deploy the
subdomain SATA on a reachable site, except possibly during certificate
issuance. It may also simplify SATisfaction
of an existing site by permitting a simple front end to handle new
authentication tasks, no change of internal links or content should be needed.
Perhaps most importantly, this SATA format is less likely to cause user
confusion. As Reynolds et al.\ recently
observed~\cite{url-identity-confusion-chi20}, ``when examining
confusing URL transforms, we found that users were least able to
understand URLs with long subdomains/FQDNs.''  There are thus usable
security advantages to adopting a query-string format for SATAs. 


\subsubsection{Authentication using a SAT address}
\label{sec:auth}

Validating the certificate received during
TLS session establishment is critical, especially
in a zero-trust architecture~\cite{401016}.
If the browser fails to identify a misissued or revoked cert,
the entire session can be compromised.

If the server's address is a SATA,
browser-side certificate validation can be  
greatly strengthened.
A SATA-aware browser will use the server's SATA
to authenticate the received TLS certificate.
For a SATA with the domain \texttt{bank.com}
authentication works as follows:
\begin{denseitemize}
\item \emph{Preparation:} 
Once every few days, a backend server at \texttt{bank.com}
uses the onion private key associated with the SATA
to sign the fingerprints of all the TLS certificates at \texttt{bank.com}
(one signature per certificate).
Let $S$ be the resulting signature data for one such certificate.

\item \emph{Server sends signature:} 
When a browser and a server at \texttt{bank.com}
establish a TLS connection, 
the server sends appropriate signature data $S$ to the browser.
This signature is either embedded in a new
HTTP header called a \emph{SATA header},
or is sent in a TLS \texttt{server-hello} extension.
Our WebExtension uses the header method because it
cannot read \texttt{server-hello} data.
This header is a form of sattestation, about which more will be
said presently.

\item \emph{Validation by browser:}
The browser parses the SATA HTTP header and 
verifies the embedded signature data using the public key
encoded in the SAT address.
If valid, it learns that the TLS certificate 
is a valid certificate for \texttt{bank.com}, as required.
Otherwise, the browser raises an error.
This is done in addition to the standard 
browser-side TLS certificate validation checks. 
\end{denseitemize}
Note that the browser need not use the public key for lookup of
and connection to an onion service. 
The browser may not even be capable
of invoking connections via the Tor network.
The browser uses the public key embedded in the
SATA only to verify a signature.

This SATA validation process ensures that an attacker who wants to 
masquerade as \texttt{bank.com}
must obtain a fake certificate for\\ \texttt{bank.com}
\emph{and} 
must obtain a valid SATA header for that fake certificate
signed by the private onion key for this SATA
for \texttt{bank.com}.

\medskip
A few more comments about this approach. 
First, the SATA for \texttt{bank.com} should be embedded in the
\texttt{bank.com} TLS certificate.  
The TLS certificate thus authenticates the SATA, and vice versa.
This also ensures that the SATA for \texttt{bank.com} is included
in CT logs that list this TLS certificate, enabling \texttt{bank.com}
to detect invalid SATAs.
Second, because the backend server used in the preparation step
is mostly offline, access to it can be more tightly controlled.

Third, the signature data $S$ sent to the browser 
in the SATA header needs to be specified more precisely.
This signature data is a special case of a new credential
format, sattestation, that is presented
in Section~\ref{sec:sattestation}.
This data contains the following fields:
\begin{itemize}[nosep]
\item version number,
\item issued date,
\item refresh rate (e.g., seven days),
\item domain name,
\item TLS certificate fingerprint(s), and
\item optionally, a comma-separated list of labels,
\end{itemize}
along with an ed25519 signature on these data fields.
The resulting data is less than 800 bytes, without compression.

Our Webextension running on the client checks if the attempted
connection is to a SATA\@. 
If not, it does nothing and returns control to the browser. 
If the URL is a SATA, it makes sure that the SATA is present in the
TLS certificate from the server. 
It then validates the SATA header by verifying that 
(i) a valid signature is present, 
(ii) that it is currently valid, 
namely that the absolute value of the 
difference between the current date and the issued date
is less than the refresh rate, and
(iii) that one of the certificate fingerprints in the SATA header 
matches the TLS certificate from the server. 
If all checks succeed, then the connection completes.
Otherwise an appropriate error message is produced.


\subsection{Sattestation}
\label{sec:sattestation}

A \emph{sattestation} is a new credential format,
defined in this paper, 
that is designed for relatively short-lived assertions by
sites about themselves (\emph{self-sattestations}), 
as well as for contextual trust assertions
by third parties that
scale both up and down.
In simplest form, a sattestation 
binds two values $D$ and $O$ to each other, where:
\begin{tightlist}
\item $D$ is a domain name (e.g., {\tt nytimes.com}), and   
\item $O$ is a self-authenticating name, such as a hashed public key
or an onion address.
\end{tightlist}
Sattestations also have two optional values
\begin{tightlist}
\item $L$ is a set of labels (e.g., {\tt \{news\}}), and
\item $C$ is a hash of a TLS Certificate.
\end{tightlist}
A sattestation is an assertion issued by a \emph{sattestor} indicating  
that $D$, $O$, (and possibly $L$, $C$) are bound to each other. 
Labels are used to reason about the trust properties 
of the sattestation.
A sattestor can be as large as a CA
that generates sattestations for the public,
 as small as an individual who is sending an
attested SATA to a friend,
or anything in between.

A single sattestation can include multiple bindings
such as\\ $((D_1,O_1), \ldots, (D_n, O_n), L)$,
indicating that $O_i$ is a self-authenticating name for $D_i$, 
for all $i=1,\ldots, n$.
This lets a sattestor issue a single sattestation for multiple
related domains (e.g., multiple news sites or multiple domains
owned by the same company). 

Fig.~\ref{fig:sattestation} gives an example sattestation
in JSON format that can be sent as a credential in a header.
The sattestation includes the sattestor's (issuer's) SATA,
the refresh rate indicating how often the sattestor verifies
the bindings and reissues the sattestation,
and the list of domain-address bindings.
Each binding indicates the date it was first issued,
and when it was last refreshed. 
Finally, the sattestation is signed by the sattestor. 
Notice that the sattestor is identified by its SATA
(in the \texttt{sattestor\_domain} and \texttt{sattestor\_onion} fields),
and not only by its domain. 
This sattestation does not bind any TLS certificate,
typical for a third-party sattestation.

\begin{figure}[t]
\begin{center}
\small
\begin{verbatim}
{ "sattestation":  {
    "sattestation_version":1,
    "sattestor_domain":"sattestora.info",
    "sattestor_onion":"..." // sattestor's addr.
    "sattestor_refresh_rate":"7 days",
    "sattestees": [
    {
    // bind domain to a self auth. address
      "domain": "domain1.info",
      "onion": "...",   // onion address
      "labels": "news",
      "issued": "2020-06-01",
      "refreshed_on": "2020-08-25"
    },
    {
    // bind domain to a self auth. adress
      "domain": "domain2.info",
      "onion": "...",   // onion address
      "labels": "union",
      "issued": "2020-06-01",
      "refreshed_on": "2020-08-25"
    }  ]  },
  // signature by sattestor
  "signature": "..." }
\end{verbatim}
\caption{\small An example sattestation in JSON format}   \label{fig:sattestation}
\end{center}
\end{figure}

The SATA header described in
Section~\ref{sec:auth} is an example of a self-sattestation,
thus it does bind a TLS certificate. 
Self-sattestations can only include a single $(D,O)$ binding,
but may bind multiple certificates for the domain.
For example, in a self-sattestation, 
the sattestees part of Fig.~\ref{fig:sattestation} (below) could look like:
\begin{verbatim}
    "sattestees": [  {   // only one entry
      "domain": "sattestora.info",
      "onion": "...",   // same as sattestor
      "cert_fingerprint": ["632B119944 ...", 
                            "23964A1368 ..."],
      "issued": "2020-06-01",
      "refreshed_on": "2020-08-25"  } ]
\end{verbatim}
Because a single signature certifies multiple certificates,
this will reduce the work to generate self-sattestations for a domain
that has a large number of certificates (say, when every machine
in the data center has a different certificate).

Self-sattestations are relatively short-lived (e.g., 3.5 days).  For
this reason, and unlike certificates, there is no need for
accompanying revocation information, such as a CRL distribution point,
or an OCSP responder address.

\medskip
\paragraph{Rotation of self-authentication keys}
If a new self-authentication key is chosen for a SATA, sattestation
permits a simple means to rotate a SATA: like rotation of PGP
keys, that a new key has been adopted can be signified by having
\emph{oldSATA} and \emph{newSATA} provide sattestations for each
other. Anyone visiting a SATA site that uses the old address can be
redirected to \emph{newSATA}\@. Provided the mutual authentication
checks are passed, this can occur automatically as can the propagation
of trust to the new address without any user interaction.
Updates for browser bookmarks can be handled similar to
any change of domain, although here the WebExtension
can rely on the persistence of the domain name if it is only the
self-authenticating portion of the address that is changing to, e.g.,
suggest a change to bookmarks. Search engines and links in websites
will also need updating, but these should have no special issues versus 
existing URL updates. 

It is important that the new SATA provide sattestation for the
old SATA, not merely the reverse, to avoid framing attacks. 
How long to maintain such rotation information is a matter of
individual policy, which can vary depending on circumstances.
Once the old SATA is considered expired, it should
only be maintained with a self-sattestation that has a single
label \sattestor{(\{ \emph{newSATA} \})} to support redirection
and validation for clients with dated information.

An attacker that has compromised a SATA's onion key and
is capable of obtaining a TLS cert could perform such a rotation
itself. Nothing in the above would prevent that or detect any
misbehavior. Here, too, sattestation plays an important role. Sattestations
for \emph{oldSATA} from other sattestors trusted by a client will
not automatically propagate to \emph{newSATA}. The attacker would
also have to trick a trusted sattestor to attest to \emph{newSATA}
for the attack to succeed.

\medskip
\paragraph{Implementing Contextual Trust}
Previous work mentioned the idea of
sattestation~\cite{secdev19} but had only a toy representation
and implementation via
downloadable lists of sattestee addresses: unlike the sattestations
introduced in this paper there were no sattestation
credential mechanisms, no means to cope with any transitivity (as
permitted by a \sattestor{} label),
and no means to validate or even describe labels or context. Thus it
had no means to provide sattestations for a particular corporation,
industry group, societal category, government agency, etc.

\subsection{Implementation}
\label{sec:implementation}

We developed a prototype implementation of our contributions by
extending the prototype Firefox WebExtension already publically
available~\cite{secdev19}. 
The extension is available for download\footnote{
  \texttt{https://selfauthdomain.info/sata.xpi?onion=hmfakusaluamq46u43uncnedm 
    cpsjks46en77mzbzb32x3en3gpkh3ad}\\ or at
  \texttt{https://github.com/sysrqb/satis-selfauth-domains}.}.

Our implementation adds support for
query-string formatted SATAs, creation of sattestation
credentials at a well-known address on a server, and verification
on a client of valid sattested SATAs via credentials. We
re-implemented the server-side component in Rust, allowing for easier
testing and deployment than would be possible by directly modifying
the Tor source code.  This server-side program produces SATA
headers and sattestations. Then an Apache web server configuration
is modified so the correct HTTP headers are sent to clients.
We also implemented a custom serialization format for sattestation
lists and credentials due to the lack of an existing web standard for
cryptographic signatures using ed25519.

Tor's onion services do not rely on DNS for address lookup. Instead a
DHT comprised of thousands of Tor relays stores network location
information to access onion services. Descriptors with this and other
information are encrypted so that, unless a DHT relay already knows
the onion address, it will not know for what onion service it is
holding descriptors. Descriptors are also regularly reassigned to
unpredictable nodes within the DHT\@. For these and other reasons, it
would thus be advantageous when Tor Browser attempts to connect to a
SATA if lookup and connection uses onion service protocols. We have
now implemented this.

Also, 
a SATA in the URL bar
prompting such an onion service connection is not rewritten to the
corresponding .onion address. Amongst other things, this means that
sites with SATAs can now easily have a certified TLS connection when
contacted in this way via Tor Browser. Though DV certificates for
.onion addresses have been approved by CA/Browser Forum guidelines
since March 02020, almost all CAs have only issued EV certificates for .onion
addresses to date. (Recently HARICA became the only exception.) Since
SATAs can have DV certificates, with our implementation the security
advantages of onion services, SATAs, and TLS certificates can all
occur together for connections to SATAs from Tor Browser.

\end{document}